\documentclass[aps,prd,twocolumn,showpacs,preprintnumbers,amsmath,amssymb]{revtex4-1}
\usepackage{graphicx}
\usepackage[usenames]{color}
\usepackage{amsmath}
\usepackage{float}
\usepackage{graphicx}
\usepackage{epstopdf}
\usepackage{amssymb}
\usepackage{extarrows}
\usepackage{color}
\newcommand{\be}{\begin{equation}}
\newcommand{\ee}{\end{equation}}
\newcommand{\gr}{\mathrm{GR}}
\newcommand{\gw}{\mathrm{GW}}

\newcommand{\h}{\mathrm{Hubble}}
\newcommand{\ligo}{\mathrm{LIGO}}
\newcommand{\lisa}{\mathrm{LISA}}
\newcommand{\m}{\mathrm{max}}

\newcommand{\oct}{\mathrm{Oct}}

\newcommand{\peak}{\mathrm{peak}}

\newcommand{\au}{\mathrm{AU}}

\def\e1{e_1^2}
\def\I{i_{\mathrm{tot}}}

\begin{document}

\preprint{APS}

\title{Merging compact binaries in hierarchical triple systems:
  Resonant excitation of binary eccentricity}

%\thanks{Electronic address: dong@astro.cornell.edu }

%DL2: add/change below
\author{Bin Liu$^{1,2,3}$}
\author{Dong Lai$^{3}$}
%\email{dong@astro.cornell.edu}
\author{Ye-Fei Yuan$^{1,2}$}

\affiliation{$^1$ Department of Astronomy, University of Science and Technology of China, Hefei, Anhui 230026, China}
\affiliation{$^2$ CAS Key Laboratory for Research in Galaxies and Cosmology, Hefei, Anhui 230026, China}
\affiliation{$^3$ Cornell Center for Astrophysics and Planetary Science,
Cornell University, Ithaca, New York 14853, USA}

\date{\today}
\begin{abstract}
%DL3: I have rewritten it
We study the secular dynamics of compact binaries (consisting of white
dwarfs, neutron stars or black holes) with tertiary companions in
hierarchical triple systems.  As the inner binary (with initially
negligible eccentricity) undergoes orbital decay due to gravitational
radiation, its eccentricity can be excited by gravitational forcing
from the tertiary. This excitation occurs when the triple
system passes through an ``apsidal precession resonance", when the
precession rate of the inner binary, driven by the gravitational
perturbation of the external companion and general relativity, matches
the precession rate of the outer binary. The eccentricity excitation
requires the outer companion to be on an eccentric orbit, with the
mutual inclination between the inner and outer orbits less than $\sim
40^\circ$. Gravitational wave (GW) signals from the inner binary
can be significantly modified as the system evolves through the
apsidal precession resonance. For some system parameters (e.g., a white dwarf
binary with a brown dwarf tertiary), the resonance can happen when the
binary emits GWs in the $10^{-4}$-$10^{-1}$~Hz range (the sensitivity
band of LISA).
\end{abstract}

\maketitle

\section{Introduction}

%DL2:rewrite
Merging compact binaries containing white dwarfs (WDs), neutron star
(NSs) and black holes (BHs) are of great importance and current
interest in astronomy.  First, depending on their total mass, WD-WD
binaries may evolve into type Ia supernova, NS formation through
accretion-induced collapse, or AM CVn binaries or R CrB stars (e.g.,
\cite{Webbink}, \cite{IT}); merging NS-NS or NS-BH binaries have been
conceived as the progenitors of short gamma-ray bursts (GRBs) (e.g.,
\cite{Eichler}, \cite{Berger}, \cite{Nissanke}).  Second, merging
compact binaries are promising sources of gravitational waves
(GWs), detectable by the Laser Interferometer Space Antenna (LISA) (see \cite{Nelemans}, \cite{elisascience})
%DL2: Above, pls add: https://www.elisascience.org/
when the the corresponding GW frequencies are in the
$10^{-4}$-$10^{-1}$ Hz band, or by the ground-based interferometers
(LIGO/VIRGO) in the next decade (see \citep{LIGO}) when the GW
frequencies lie in the 10-1000 Hz band.  For detecting these GWs,
accurate gravitational waveforms during the binary inspiral and merger
phases are needed, and the latter requires three-dimensional
numerical simulations in
full general relativity (e.g., \cite{Cutler}, \cite{Shibata},
\cite{Foucart}, \cite{Sekiguchi}).

%Merging compact binaries, such as white draft (WD) binaries, neutron
%star (NS) binaries and black hole (BH) binaries, are of great
%importance and current interest in astronomy.  First, depending on
%their total mass, WD-WD binaries may evolve into Type Ia supernova
%(high mass) or AM CVn binaries or R CrB starts (low mass) (e.g.,
%\cite{Webbink}, \cite{IT}); NS-NS or NS-BH binaries have been
%conceived as the progenitors of short gamma ray bursts (GRBs) (e.g.,
%\cite{Eichler}, \cite{Berger}, \cite{Nissanke}).  Second, merging
%binaries are suggested to be promising sources of gravitational waves
%(GWs), and the corresponding frequencies are in the
%$10^{-4}$-$10^{-1}$ Hz band that are detectable by the Laser
%Interferometer Space Antenna (LISA) (see \cite{Nelemans}) or in the
%10-1000 Hz band that will be potentially detected by ground-based
%interferometers (LIGO/VIRGO) in the next few years (see \citep{LIGO}).
%For the requirement of detection of GWs, recently, the
%inspiral-merger-ringdown waveforms of the mergers are obtained through
%3D numerical relativity simulations (e.g., \cite{Cutler},
%\cite{Shibata}, \cite{Foucart}, \cite{Sekiguchi}).

In this paper, we study the evolution of merging compact binaries
influenced by an external third body.  Three-body systems are
ubiquitous in astrophysics, appearing in a wide range of
configurations and scales. A significant fraction of stars are in triples
\cite{Tokovinin}.
%DL2: above, please add ref: http://adsabs.harvard.edu/abs/2014AJ....147...87T
%DL2: add below
It is possible that a tertiary companion orbits around the compact
binary as part of the original stellar triple. A low-mass companion
may also form in the envelope around the compact binary (during
the common envelope phase of binary evolution). In such systems,
long-term secular interaction between the inner and outer orbits may
induce variations of the eccentricity and inclination of the inner
binary. In particular, when the mutual inclination of two orbits lies between
$40^\circ$ and $140^\circ$, the eccentricity and inclination may undergo
oscillations -- the so-called Lidov-Kozai oscillations
(e.g., \cite{Lidov}, \cite{Kozai}).
The effects of Lidov-Kozai oscillations in the formation and evolution of various
astrophysical systems have been extensively studied in recent years
(e.g., \cite{Eggleton}, \cite{FT},
\cite{Holman}, \cite{Innanen}, \cite{WM}, \cite{Smadar 2011},
\cite{Thompson}, \cite{PMT}, \cite{Katz PRL}, \cite{Blaes}, \cite{MH},
\cite{Wen}, \cite{AMM}, \cite{DongLai1}, \cite{DongLai2}).
%DL2: Above, please refs:
%http://adsabs.harvard.edu/abs/2014Sci...345.1317S
%http://adsabs.harvard.edu/abs/2015MNRAS.448.1821S
It has also been recognized that high-order (octupole) interactions
between the inner and outer orbits can lead to rich dynamical
behaviors in hierarchical triples (e.g., \cite{Harrington},
\cite{Marchal}, \cite{Ford}, \cite{Smadar 2013b}, \cite{SRF}).

%DL2:
In this paper we focus on systems with small mutual inclinations so
that Lidov-Kozai oscillations do not occur.  Several recent studies
(e.g. \cite{Ford}, \cite{Mardling 2007} and \cite{Smadar 2013a})
showed that when post-Newtonian effect of general relativity (GR) is
taken into account in the secular dynamics, significant eccentricity
excitation can still occur under some conditions.
In particular, periapse (apsidal) precession of the inner orbit
due to GR plays an important role. However, these studies did not identify
the physical origin and the condition of eccentricity excitation.

Here we show that eccentricity excitation arises from a resonance
between the precessions of the inner and outer binaries due to the
combined effects of mutual gravitational interaction and GR. In
essence, the inner binary ``gets'' its eccentricity from the outer
binary through this ``apsidal precession resonance".  We also study
how the eccentricity of the inner binary evolves as the system crosses
the resonance driven by GW-induced orbital decay of the inner binary.

Our paper is organized as follows.  In Sec. II, we present the
secular evolution equations for coplanar triple systems.  In Sec.III,
a linear eigenmode analysis for the eccentricities of the inner and
outer binaries is given; the result is used to identify the ``apsidal
precession resonance'' and to calculate the maximum eccentricity that
can be achieved in the resonance.  In Sec. IV, we explore the
parameter space for resonances in the various systems.  In Sec. V,
we describe our numerical integrations for the evolution of the triple
when the inner binary undergoes orbital decay due to GW emission.  In
Sec. VI, we extend our calculations to systems with arbitrary mutual
inclinations. Our results and conclusions are summarized in Sec. VII.

%%%%%%%%%%%%%%%%%%%%%%%%%%%%%%%%%%%%%%
\section{Secular Dynamics of Stellar Triples: Equations}

\subsection{Disturbing function}

\begin{figure}
\centering
\begin{tabular}{cc}
\includegraphics[width=7cm]{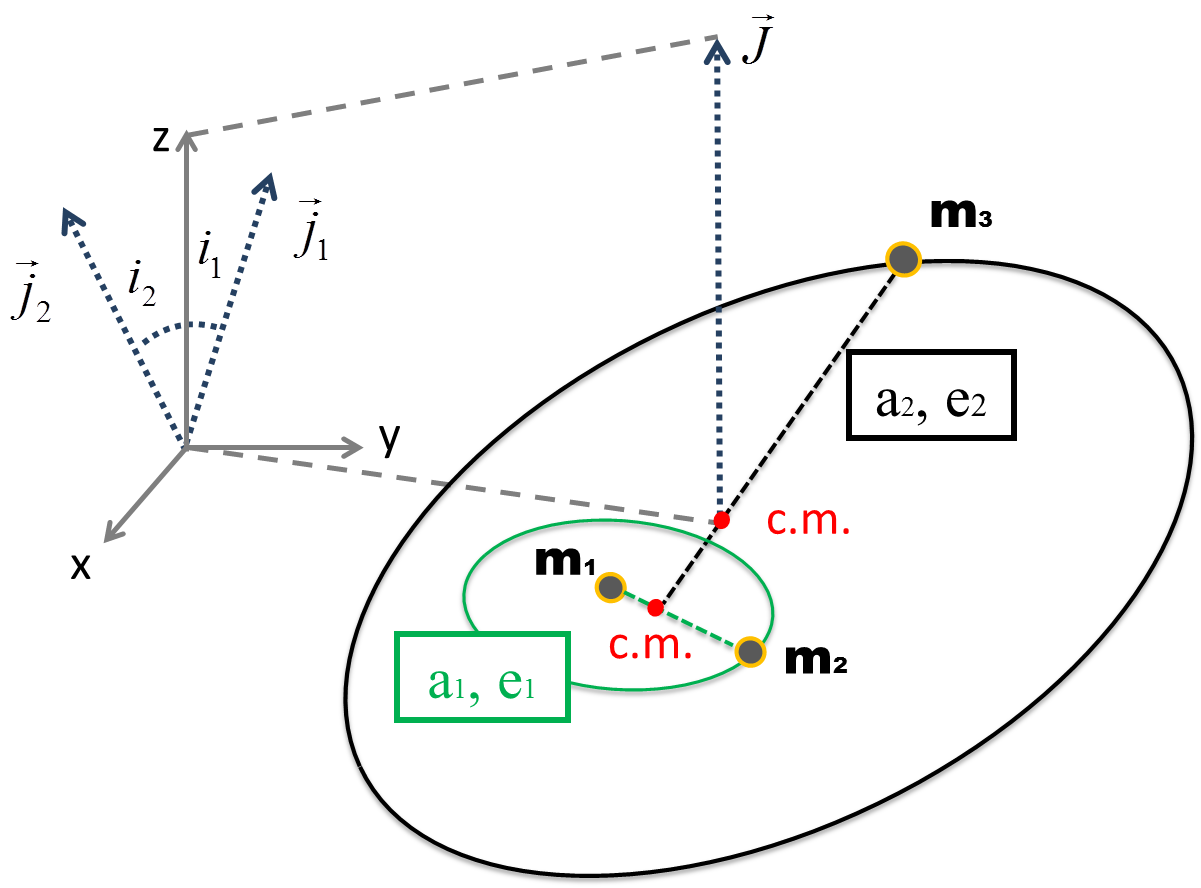}
\end{tabular}
%DL2: modified
\caption{Coordinate system used to describe the hierarchical triple system,
where $z$-axis is aligned with the total angular momentum $\textbf{J}$.
Here, $\textbf{j}_1$ and $\textbf{j}_2$ denote the angular momenta of the inner and outer binaries, respectively,
while ``c.m." indicates the center of mass of each system.
}\label{fig:1}
\end{figure}

We consider the evolution of a hierarchical triple system, consisting of
an inner binary with the masses of $m_1$, $m_2$
and a relatively distant third body of mass $m_3$.
We introduce the orbital semimajor axes as $a_{1,2}$ and eccentricities as $e_{1,2}$,
the subscripts ``$1,2$" denote the inner and outer binaries, respectively.
%DL:
%The position of
Each orbit is described by three angles: the inclination $i$, the
longitude of the periapse $\omega$ and the longitude of the ascending
node $\Omega$, which are defined in a coordinate where the $z$ axis is
aligned with the total angular momentum.  Thus, the relative
inclination between the two orbits is $\I\equiv i_1+i_2$, as shown in Fig. \ref{fig:1}.

Based on the calculation in \cite{Ford}, \cite{Smadar 2013b} and \cite{SRF},
%DL:
the interaction potential, averaged over both the inner and outer orbits and
expanded up to the octupole order,
is given by
\be\label{eq:Potential}
\begin{split}
\langle\langle\Phi\rangle\rangle&=-\frac{\mu_1\Phi_0}{8}\Big[2+3\e1-
(3+12\e1-15\e1\cos^2\omega_1)\sin^2\I\Big]\\
&+\frac{15\mu_1\Phi_0\varepsilon_{\oct}}{64}\Bigg\{5e_1(1-\e1)
\sin\I\sin2\I\sin\omega_1\\
&\times\sin\omega_2+\frac{e_1}{4}\Big[6-13\e1+5(2+5\e1)
\cos2\I\\
&+70\e1\cos2\omega_1\sin^2\I\Big]\\
&\times\Big(-\cos\omega_1\cos\omega_2-\cos\I\sin\omega_1\sin\omega_2\Big)\Bigg\}~,
\end{split}
\ee
where
%DL:
\be
\begin{split}
&\mu_1=\frac{m_1m_2}{m_1+m_2},\\
&\mu_2=\frac{(m_1+m_2)m_3}{m_1+m_2+m_3}
\end{split}
\ee
are the reduced masses of the inner and outer orbits, respectively.
In Eq.(\ref{eq:Potential}), the coefficients are defined as:
\be
\Phi_0\equiv\frac{Gm_3a_1^2}{a_2^3(1-e_2^2)^{3/2}},~
\ee
where $G$ is the gravitational constant and
\begin{equation}\label{eq:C}
\varepsilon_{\oct}\equiv\frac{m_1-m_2}{m_1+m_2}\frac{a_1}{a_2}\frac{e_2}{1-e_2^2}
\end{equation}
%DL:
quantifies the relative significance of the octupole term.

\subsection{Equations of motion with GR effects}

%DL
Various short-range effects, such as general relativity (GR), tidal and
rotational distortions, can
%DL:
affect the secular dynamics or the stellar triples
%limit the growth of eccentricity and
%suppress the Lidov-Kozai oscillations
(e.g., \cite{FT}, \cite{SRF}).
Since the tidal and rotational effects on the compact objects are small compared to the GR effect,
%DL:
here we only consider the leading order GR effect on the inner binary
%study the dynamics of triple systems including the GR effects but
in the absence of energy dissipation
%DL:
due to gravitational waves (see Sec. III).
%DL: Pls add above
Thus, the total potential   $\langle\langle\Phi_{\mathrm{tot}}\rangle\rangle=\langle\langle\Phi\rangle\rangle+\langle\langle\Phi_{\gr}\rangle\rangle$ is conserved,
where
\be\label{eq:PGR}
\langle\langle\Phi_{\gr}\rangle\rangle=-\frac{3G^2m_1m_2(m_1+m_2)}{a_1^2c^2}\frac{1}{(1-\e1)^{1/2}}
\ee
is the post-Newtonian potential associated with the
%DL:
GR-induced pericenter advance (e.g., \cite{Eggleton}).

%DL: Pls add XXX below
Our numerical calculations (see Sec. \ref{chap:inclined})
will consider general inclinations between the inner and outer binaries.
Here we focus on coplanar configurations, with $\I =0$.
We define
% use $\varpi_{1,2}$ instead of $\omega_{1,2}$, which are satisfied with the relation:
%
\be\label{eq:varpi}
\varpi_1\equiv\omega_1+\Omega_1,~~~~~~
\varpi_2\equiv\omega_2+\Omega_2.
\ee
Substituting Eq.(\ref{eq:varpi}) into Eq.(\ref{eq:Potential}) and
Eq.(\ref{eq:PGR}),
%DL:
the total potential can only be expressed in terms
of $e_{1,2}$ and $\varpi_{1,2}$.  The orbital elements
are governed by the following Lagrange's equations \cite{MD}:
\be\label{eq:H eqs1}
\mu_1\dot{e_1}=\frac{\sqrt{1-e_1^2}}{n_1 a_1^2 e_1}\frac{\partial\langle\langle\Phi_{\mathrm{tot}}\rangle\rangle}{\partial\varpi_1},
\ee
\be\label{eq:H eqs2}
\mu_2\dot{e_2}=\frac{\sqrt{1-e_2^2}}{n_2 a_2^2 e_2}\frac{\partial\langle\langle\Phi_{\mathrm{tot}}\rangle\rangle}{\partial\varpi_2},
\ee
\be\label{eq:H eqs3}
\mu_1\dot{\varpi_1}=-\frac{\sqrt{1-e_1^2}}{n_1 a_1^2 e_1}\frac{\partial\langle\langle\Phi_{\mathrm{tot}}\rangle\rangle}{\partial e_1}
-\frac{\tan(\I/2)}{n_1 a_1^2\sqrt{1-e_1^2}}\frac{\partial\langle\langle\Phi_{\mathrm{tot}}\rangle\rangle}{\partial \I},
\ee
\be\label{eq:H eqs4}
\mu_2\dot{\varpi_2}=-\frac{\sqrt{1-e_2^2}}{n_2 a_2^2 e_2}\frac{\partial\langle\langle\Phi_{\mathrm{tot}}\rangle\rangle}{\partial e_2}
-\frac{\tan(\I/2)}{n_2 a_2^2\sqrt{1-e_2^2}}\frac{\partial\langle\langle\Phi_{\mathrm{tot}}\rangle\rangle}{\partial \I},
\ee
where $n_1=G^{1/2}(m_1+m_2)^{1/2}/a_1^{3/2}$ and $n_2=G^{1/2}(m_1+m_2+m_3)^{1/2}/a_2^{3/2}$
are the mean motions of the inner and outer binaries.
Thus, the evolution equations for the eccentricities (to the octupole order) are
\be\label{eq:C eqs1}
\begin{split}
\dot e_1&=-\frac{15}{64}n_1e_2\frac{\sqrt{1-e_1^2}(4+3e_1^2)}{(1-e_2^2)^{5/2}}\frac{m_3(m_1-m_2)}{(m_1+m_2)^2}\\
&\times\bigg(\frac{a_1}{a_2}\bigg)^4\sin(\varpi_1-\varpi_2),
\end{split}
\ee
and
\be\label{eq:C eqs2}
\begin{split}
\dot e_2&=\frac{15}{64}n_2e_1\frac{(4+3e_1^2)}{(1-e_2^2)^{2}}\frac{m_1m_2(m_1-m_2)}{(m_1+m_2)^3}\\
&\times\bigg(\frac{a_1}{a_2}\bigg)^3\sin(\varpi_1-\varpi_2).
\end{split}
\ee
The argument of periapse for the inner and outer binaries evolve according to
\be\label{eq:C eqs3}
\begin{split}
\dot \varpi_1&=\frac{3}{4}\bigg(\frac{a_1}{a_2}\bigg)^3n_1\frac{m_3}{m_1+m_2}\frac{\sqrt{1-e_1^2}}{(1-e_2^2)^{3/2}}\bigg[1-\\
&\frac{5a_1e_2(4+9e_1^2)(m_1-m_2)\cos(\varpi_1-\varpi_2)}{16a_2e_1(1-e_2^2)(m_1+m_2)}\bigg]+\dot\omega_\gr,
\end{split}
\ee
and
\be\label{eq:C eqs4}
\begin{split}
\dot \varpi_2&=\frac{3}{8}\bigg(\frac{a_1}{a_2}\bigg)^2n_2\frac{m_1 m_2}{(m_1+m_2)^2}\frac{2+3e_1^2}{(1-e_2^2)^{2}}\bigg[1-
\frac{5a_1e_1}{8a_2e_2}\\
&\times\frac{(4+3e_1^2)(1+4e_2^2)(m_1-m_2)\cos(\varpi_1-\varpi_2)}{(2+3e_1^2)(1-e_2^2)(m_1+m_2)}\bigg].
\end{split}
\ee
Note that the first term of Eqs.(\ref{eq:C eqs3})-(\ref{eq:C eqs4}) corresponds to the quadrupole potential
and the second term is induced by the octupole effects.
Our results agree with \cite{Mardling 2007} in the limit of small eccentricity and small $m_2$ of the inner binary.

The only effect of GR potential [Eq.(\ref{eq:PGR})] is to provide an
additional precession rate of the inner binary (e.g., \cite{FT}):
\be\label{eq:omegagr}
\dot\omega_\gr=\frac{3n_1}{1-\e1}\frac{G(m_1+m_2)}{a_1c^2}.
\ee
%
%DL:
Therefore, Eq.(\ref{eq:C eqs3}) describes the pericenter precession of
the inner binary due to the combined effects of GR and the external body
$m_3$.

\section{Linear Analysis: Apsidal Precession Resonance}
\label{chap:Linear}

\subsection{Analytical equations}

For $e_1, e_2 \ll 1$, Eqs.(\ref{eq:C eqs1})-(\ref{eq:C eqs4}) can be reduced to
\be\label{eq:our eqs1}
\dot e_1=-\frac{15}{16}n_1e_2\frac{m_3(m_1-m_2)}{(m_1+m_2)^2}
\bigg(\frac{a_1}{a_2}\bigg)^4\sin(\varpi_1-\varpi_2),
\ee
\be\label{eq:our eqs2}
\dot e_2=\frac{15}{16}n_2e_1\frac{m_1m_2(m_1-m_2)}{(m_1+m_2)^3}
\bigg(\frac{a_1}{a_2}\bigg)^3\sin(\varpi_1-\varpi_2),
\ee
\be\label{eq:our eqs3}
\begin{split}
\dot \varpi_1&=\frac{3}{4}\bigg(\frac{a_1}{a_2}\bigg)^3n_1\frac{m_3}{m_1+m_2}\\
&\times\bigg[1-\frac{5}{4}\frac{a_1}{a_2}\frac{e_2}{e_1}\frac{m_1-m_2}{m_1+m_2}\cos(\varpi_1-\varpi_2)\bigg]+\dot\varpi_\gr,
\end{split}
\ee
\be\label{eq:our eqs4}
\begin{split}
\dot \varpi_2&=\frac{3}{4}\bigg(\frac{a_1}{a_2}\bigg)^2n_2\frac{m_1m_2}{(m_1+m_2)^2}\\
&\times\bigg[1-\frac{5}{4}\frac{a_1}{a_2}\frac{e_1}{e_2}\frac{m_1-m_2}{m_1+m_2}\cos(\varpi_1-\varpi_2)\bigg].
\end{split}
\ee
Following \cite{MD}, we introduce the complex variable
$I_\alpha\equiv e_\alpha e^{i\varpi_\alpha}$ (with $\alpha=1,2$).  Then
Eqs.(\ref{eq:our eqs1})-(\ref{eq:our eqs4}) can be combined to give:
\be\label{eq:I1I2}
\dot I_1=iA_{11}I_1+iA_{12}I_2,~~~~~~\dot I_2=iA_{21}I_2+iA_{22}I_2.
\ee
where the coefficients $A_{ij}$ are given by
\begin{eqnarray}
&A_{11}=\frac{3}{4}n_1\frac{m_3}{m_1+m_2}\bigg(\frac{a_1}{a_2}\bigg)^3+\dot\omega_\gr|_{e_1\rightarrow0},\label{eq:A11}\\
&A_{12}=-\frac{15}{16}n_1\bigg(\frac{a_1}{a_2}\bigg)^4\frac{m_3(m_1-m_2)}{(m_1+m_2)^2},\label{eq:A12}\\
&A_{21}=-\frac{15}{16}n_2\bigg(\frac{a_1}{a_2}\bigg)^3\frac{m_1m_2(m_1-m_2)}{(m_1+m_2)^3},\label{eq:A21}\\
&A_{22}=\frac{3}{4}n_2\frac{m_1m_2}{(m_1+m_2)^2}\bigg(\frac{a_1}{a_2}\bigg)^2. \label{eq:A22}
\end{eqnarray}
Clearly, $A_{11}$ is the total apsidal precession rate of the inner
binary induced by the outer companion (to the quadrupole order) and
the GR corrections, while $A_{22}$ is the precession rate of the outer
binary driven by the quadrupole potential of the inner binary.

The standard Eq.(\ref{eq:I1I2}) can also be found in \cite{Wu G} in the
case of $m_2, m_3\ll m_1$ but without requiring $a_1/a_2\ll 1$.
Setting $I_\alpha \sim e^{i g_\star t}$, we obtain the two
eigenvalues:
\be\label{eq:g12}
g_{\star1,2}=\frac{A_{11}+A_{22}\mp\sqrt{(A_{11}-A_{22})^2+4A_{12}A_{21}}}{2},
\ee
where $g_{\star 1}$ refers to the upper sign, and $g_{\star 2}$ refers
to the lower sign. The corresponding linear eigenvectors are:
\be\label{eq:xi}
\xi_1=
\begin{pmatrix}
\xi_{11}\\ \xi_{12}
\end{pmatrix}=
\begin{pmatrix}
\frac{g_{\ast1}-A_{22}}{A_{21}}\\1
\end{pmatrix},~
\xi_2=
\begin{pmatrix}
\xi_{21}\\ \xi_{22}
\end{pmatrix}=
\begin{pmatrix}
\frac{g_{\ast2}-A_{22}}{A_{21}}\\1
\end{pmatrix}.
\ee

\subsection{Excitation of eccentricity due to resonance}

The general solution of Eq.(\ref{eq:I1I2}) can be written in the form
\be\label{eq:I1I22}
\begin{pmatrix}
I_1\\I_2
\end{pmatrix}=A
\begin{pmatrix}
\xi_{11}\\ \xi_{12}
\end{pmatrix}e^{ig_{\ast1}t}+B
\begin{pmatrix}
\xi_{21}\\ \xi_{22}
\end{pmatrix}e^{ig_{\ast2}t}.
\ee
The resulting time evolution of the eccentricity of the inner binary is then
\be\label{eq:E1}
e_1=\Big|(A\xi_{11})^2+(B\xi_{21})^2+2AB\xi_{11}\xi_{21}\cos[(g_{\ast1}-g_{\ast2})t]\Big|^{1/2},
\ee
and for the outer binary
\be\label{eq:E2}
e_2=\Big|(A\xi_{12})^2+(B\xi_{22})^2+2AB\xi_{12}\xi_{22}\cos[(g_{\ast1}-g_{\ast2})t]\Big|^{1/2},
\ee
where $A$ and $B$ are determined by the initial eccentricities:
\be\label{eq:initiale}
e_{1,0}=A\xi_{11}+B\xi_{21},~~~~ e_{2,0}=A+B.
\ee
Here the subscript ``0" denotes $t=0$.

Considering the case where $e_{1,0}=0$ and
%DL:
$e_{2,0}\neq0$.
%$e_2\neq0$,
%
Then $e_1(t)$ will oscillate between $0$ and $e_{1,\m}$, which is given by
\be\label{eq:E1MAX}
e_{1,\m}=|A\xi_{11}-B\xi_{21}|=2e_{2,0}\left|\frac{A_{12}}{g_{\star2}-g_{\star1}}\right|.
\ee

Clearly, $e_{1,\m}$ achieves the largest value, $e_{1,\peak}$,
when $|g_{\star1}-g_{\star2}|$ is smallest or, equivalently,
\be\label{eq:Res}
A_{11}=A_{22}.
\ee
Physically, the condition (\ref{eq:Res}) corresponds to the
$\dot\varpi_1=\dot\varpi_2$ at the quadrupole order.
The peak value in $e_{1,\m}$ at this
%DL:
{\it ``apsidal precession resonance"} is then
\be\begin{split}\label{eq:peak}
e_{1,\peak}&=e_{2,0}\bigg(\frac{A_{12}}{A_{21}}\bigg)^{1/2}=e_{2,0}\bigg(\frac{n_1}{n_2}\frac{a_1}{a_2}\frac{m_3}{\mu_1}\bigg)^{1/2}\\
&=e_{2,0}\bigg[\frac{(m_1+m_2)^3}{m_1+m_2+m_3}\bigg]^{1/4}\bigg(\frac{m_3}{m_1 m_2}\bigg)^{1/2}\bigg(\frac{a_2}{a_1}\bigg)^{1/4}.
\end{split}
\ee
Therefore, $e_{1,\peak}$ approaches $e_{2,0}$ only in the special case
of $A_{12}=A_{21}$,
%DL: I have deleted below: not sure what you mean
% Namely, upon the ``apsidal precession resonance",
%the equivalent octupole components ($A_{12}=A_{21}$) induced by the
%critical $m_3$ may even provide the highest $e_{1,\peak}$.
where the apsidal precession resonance is satisfied up to the octupole order.

\section{Parameter regime for resonances in various systems}

\subsection{Timescales}

%DL: rewrite
We are interested in inner binaries undergoing orbital decay due to
gravitational radiation. The timescale of orbital decay is given by
%For the most close binaries, two components of the inner bodies will
%merger due to the GWs radiation.  The time scale associated with the
%decay of semimajor axis is provided by
\cite{peters}
\be\begin{split}
T_\gw&\simeq3.6\times10^{12}yr\bigg(\frac{m_1}{M_\odot}\bigg)^{-1}\bigg(\frac{m_2}{M_\odot}\bigg)^{-1}\\
&\times\bigg(\frac{m_1+m_2}{M_\odot}\bigg)^{-1}\bigg(\frac{a_1}{0.05 \au}\bigg)^{4}.
\end{split}
\ee
%
%DL:
We require $T_\gw$ to be less than the Hubble time. This puts an upper
limit on $a_1$, given by
\be\label{eq:crita}
T_\gw(a_{1,\h})=T_\h\simeq1.4\times10^{10}{\rm yr}.
\ee
%DL: rewrite
For $a_1>a_{1,\h}$, the only way to have a shorter merging time
($T_\gw<T_\h$) is to rely on a tertiary companion to induce Lidov-Kozai
oscillations of the inner binary (e.g., \cite{Thompson}).
% and it is not captured in our current study.

%DL:
If resonant excitation of $e_1$ occurs
during the shrinkage of the inner binary,
the gravitational wave (GW) will be affected.
The peak GW frequency can be approximated by \cite{Wen}:
% the GWs frequency is expected to
%be modified from the fiducial predictions.  For the eccentric orbits,
%DL:
\be\label{eq:crita1}
f_\gw =\frac{(1+e_{1})^{1.1954}}{\pi}\sqrt{\frac{G(m_1+m_2)}{a_1^3(1-e_{1}^2)^3}}.
\ee
%DL: pls check
For $e_1=0$, this becomes
\be
f_\gw\sim6.34\times10^{-8} \mathrm{Hz} \bigg(\frac{a_{1}}{\au}\bigg)^{-3/2}
\bigg(\frac{m_1+m_2}{M_\odot}\bigg)^{1/2}.
\ee
%
%where the peak GWs frequency is well approximated in \cite{Wen}.
%DL:
In the following sections, we focus on the merging compact binaries
and choose the sufficiently small initial values of semimajor axis of the inner orbit $a_{1,0}$, so that the
corresponding GWs frequency at the apsidal precession
resonance lies in the LIGO or LISA bands.

\subsection{Perturbers in resonance}

\begin{table*}
 \centering
 \begin{minipage}{150mm}
  \caption{Parameters for the canonical compact binaries.
Here $a_{1,\h}$ is the semimajor axis at which the timescale of
gravitational radiation of the binary equals the Hubble time
[see Eq.(\ref{eq:crita})], $a_{1,\lisa}$ ($10^{-4}$Hz) is the semimajor axis
of the binary corresponding to GW frequency of $10^{-4}$ Hz
[see Eq.(\ref{eq:crita1})] and similarly
for $a_{1,\lisa}$ ($10^{-1}$Hz) and $a_{1,\ligo}$ (10Hz).
  }
  \begin{tabular}{@{}llrrrrlrlrlrlr@{}}
\hline
   Parameter &  ~~~System & ~$m_1$~~~~  & $m_2$~~~~ & $a_{1,\h}$~~  &
   ~$a_{1,\lisa}$($10^{-4}$Hz) & ~~$a_{1,\lisa}$($10^{-1}$Hz) & ~~$a_{1,\ligo}$($10^{1}$Hz) \\
    &   &  ~$(\mathrm{M}_\odot)$~~  & $(\mathrm{M}_\odot)$~~ &  $(\au)$~~~~~  &
   $(\au)$~~~~~~~~ &  ~~~~~~~$(\au)$ & $(\au)$~~~~~~~~ \\
 \hline
  Case 1 & ~~WD-WD &  ~0.8~~~~ & 0.4~~~~  & $1.0\times10^{-2}$ & $7.8\times10^{-3}$~~~~ & ~~~~$7.8\times10^{-5}$  & $\sim$~~~~~~~~~~~  & \\
 \\
  Case 2 & ~~~BH-BH &  ~14~~~~~ & 7~~~~~  & $8.4\times10^{-2}$ & $2.0\times10^{-2}$~~~~ & ~~~~$2.0\times10^{-4}$  & $9.5\times10^{-6}$~~~~  & \\
\hline
\end{tabular}
\end{minipage}
\end{table*}

\begin{figure*}
\centering
%\begin{tabular}{c}
\includegraphics[width=10cm]{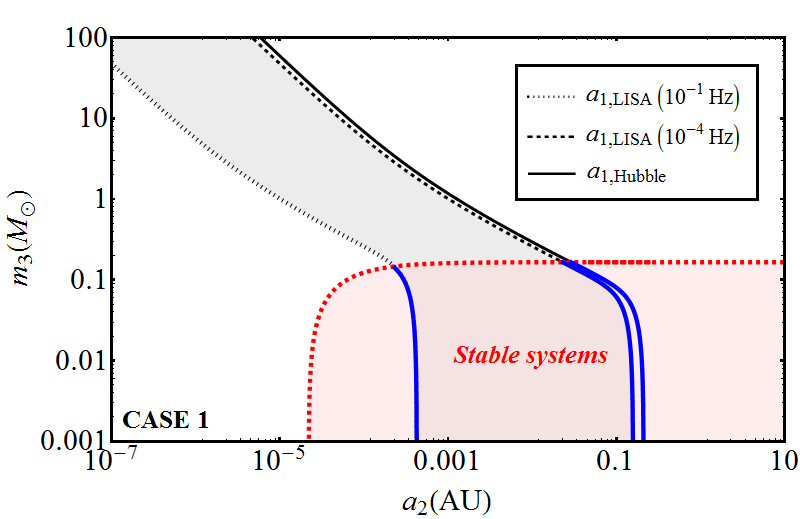}
\includegraphics[width=10cm]{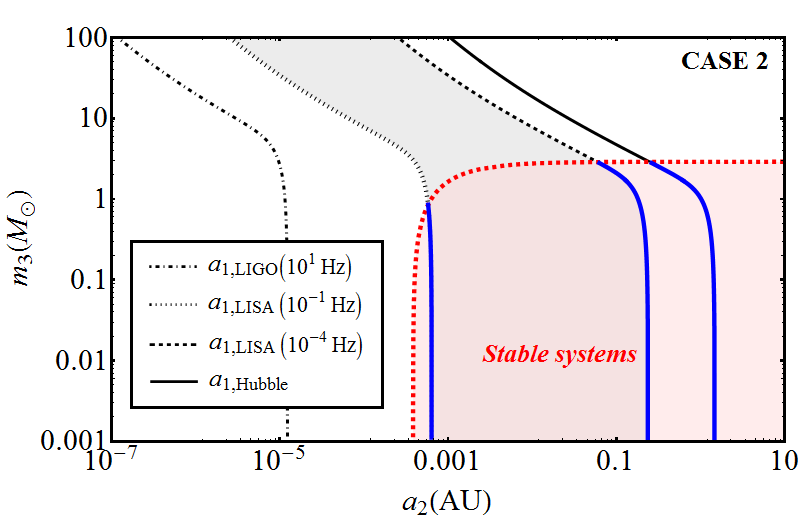}
%\end{tabular}
\caption{The mass of the third body as a function of the semimajor axis of
the outer binary when the apsidal precession resonance
occurs, for different values of semimajor axis
of the inner binary $a_1$. The upper panel shows the result for
WD-WD binaries, while the lower panel shows that for BH-BH binaries.
The masses of $m_1$ and $m_2$ are listed in Table 1.
%DL: delete
%The grey region bounded by two ``$a_{1,\lisa}$" indicates
%the parameter space for the third body where resonance is possible.
The red region corresponds to stable systems given by the criterion
[see Eq.(\ref{eq:stability})].
%DL: delete
% has further constrained the ``existence" of the perturber.
We use blue curves to mark the stable systems for different values of $a_1$.
}\label{fig:2}
\end{figure*}
%DL:
To explore the dependence of the apsidal precession resonance on the third body,
we rewrite the resonance condition (\ref{eq:Res}) as
[using Eqs.~(\ref{eq:A11}) and (\ref{eq:A22})]:
\be\label{eq:M3}
\begin{split}
m_3&=\bigg[\sqrt{\frac{m_1+m_2+m_3}{m_1+m_2}}\sqrt{\frac{a_1}{a_2}}
\frac{m_1m_2}{(m_1+m_2)^2}-4\bigg(\frac{a_1}{a_2}\bigg)^{-3}\\
&\times\frac{G(m_1+m_2)}{a_1c^2}\bigg](m_1+m_2).
\end{split}
\ee
A wide range of parameters can satisfy Eq.(\ref{eq:M3}).
Since the resonance is determined by Eq. (\ref{eq:Res})
for any mass sets regardless of the type of binaries,
we focus on two types of inner compact binaries,
with masses given in Table 1.
%If we focus on the certain regions relevant for stellar-mass compact objects,
%the physical parameters of the systems studied are listed in Table 1.

In Fig.~\ref{fig:2}, we plot $m_3$ as a function of $a_2$
for given inner binary masses ($m_1$, $m_2$) and several values of
semimajor axis $a_1$.
%DL:
%Especially, we pick out several different values of $a_1$.  So that,
Each point of the curve corresponds to a resonant triple system, i.e.,
resonance occurs if the inner binary passes through at such separation ($a_1$).
Here $a_{1,\h}$ is given by Eq.(\ref{eq:crita}), $a_{1,\lisa}$
($10^{-4}$Hz) is the semimajor axis of the binary corresponding GW
frequency entering the LISA band [see Eq.(\ref{eq:crita1})] and similarly
for $a_{1,\lisa}$ ($10^{-1}$Hz) and $a_{1,\ligo}$ $(10$Hz, LIGO band).
We see that the curves of $a_{1,\lisa}$ and $a_{1,\ligo}$ are
on the left side of curve $a_{1,\h}$.  This implies that the resonance
can happen within the Hubble time and the corresponding resonant GW
frequencies enter the LISA band or LIGO band.
% which can be detected potentially.

%DL:
However, the triple system must satisfy the stability criterion given
by \cite{MA}
\be\label{eq:stability}
\frac{a_2}{a_1}>2.8\bigg(1+\frac{m_3}{m_1+m_2}\bigg)^{2/5}\frac{(1+e_2)^{2/5}}{(1-e_2)^{6/5}}\bigg(1-\frac{0.3i}{180^\circ}\bigg).
\ee
This further constrains the ``existence" of the third body.
Thus, only those triple systems in the ``stable region" of Fig.\ref{fig:2}
are realistic.
%DL: delete
%the secular equations and linear analysis are reasonable.
%Therefore, $a_{1,\ligo}$ is not relative for the compact objects we considered
%and WD-WD binaries may be more interesting.

\begin{figure}
\centering
%\begin{tabular}{c}
\includegraphics[width=8cm]{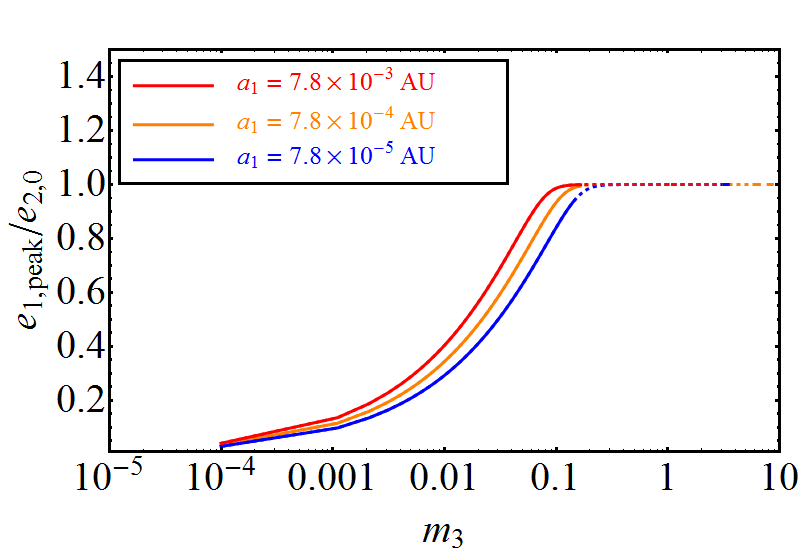}
%\end{tabular}
\caption{The peak eccentricity in units of the initial outer binary eccentricity
as a function of $m_3$ for the different values of
$a_1$.
%DL: Pls change the figure: a_1,LISA --> a_1
%$a_{1,\lisa}$.
%The three relevant ratio between the resonant eccentricity and initial
%outer eccentricity
%DL:
The dashed lines correspond to unstable systems [see Eq.(\ref{eq:stability})].
%The dashed line is given by the stability criterion (Eq.\ref{eq:stability}).
In this example, $e_{1,\peak}$ increases, approaching $e_{2,0}$,
as the external perturber mass $m_3$ increases.
}\label{fig:3}
\end{figure}

%DL:delete
%It is found that the resonance is induced by small perturber in the
%coplanar systems in Fig.\ref{fig:1}.
%DL:
Before studying resonant excitation of eccentricity
during GW-induced orbital decay, we consider a few examples to
illustrate the dependence of the peak values of the excited eccentricity on
the mass of the third body. We focus on WD-WD binaries (Case 1) and
choose three values of
%DL: In Fig.2, the label shoudl be a_1 instead of a_1,LISA. Pls change the figure
$a_1$.
%$a_{1,\lisa}$.
Combing Eqs.(\ref{eq:peak})
and (\ref{eq:M3}), setting the initial eccentricities as
$e_{1,0}=0.0001$, $e_{2,0}=0.1$, we plot the ratio
$e_{1,\peak}/e_{2,0}$ as a function of $m_3$ in Fig.~\ref{fig:3}.
We see that $e_{1,\peak}$ approaches $e_{2,0}$
as $m_3$ increases.
% but it cannot go beyond the value of $e_{2,0}$,
%which seems that $e_{2,0}$ puts an upper limit on the excitation of
%eccentricity.
%DL: delete
%The dashed line corresponds unstable systems due to
%Eq.(\ref{eq:stability}).
%As a result, based on the form of
%Eq.(\ref{eq:peak}), we find that the equivalence between two octupole
%components of periastron advancing frequency of two orbit
%($A_{12}=A_{21}$) drives the value of $e_{1,\peak}$ to be the maximum.
%In further sections, we confirm this by running a series of numerical
%integrations.

\section{Effects of the gravitational radiation}
\label{chap:coplanar GW}

\begin{figure*}
\centering
\begin{tabular}{cc}
\includegraphics[width=8.8cm]{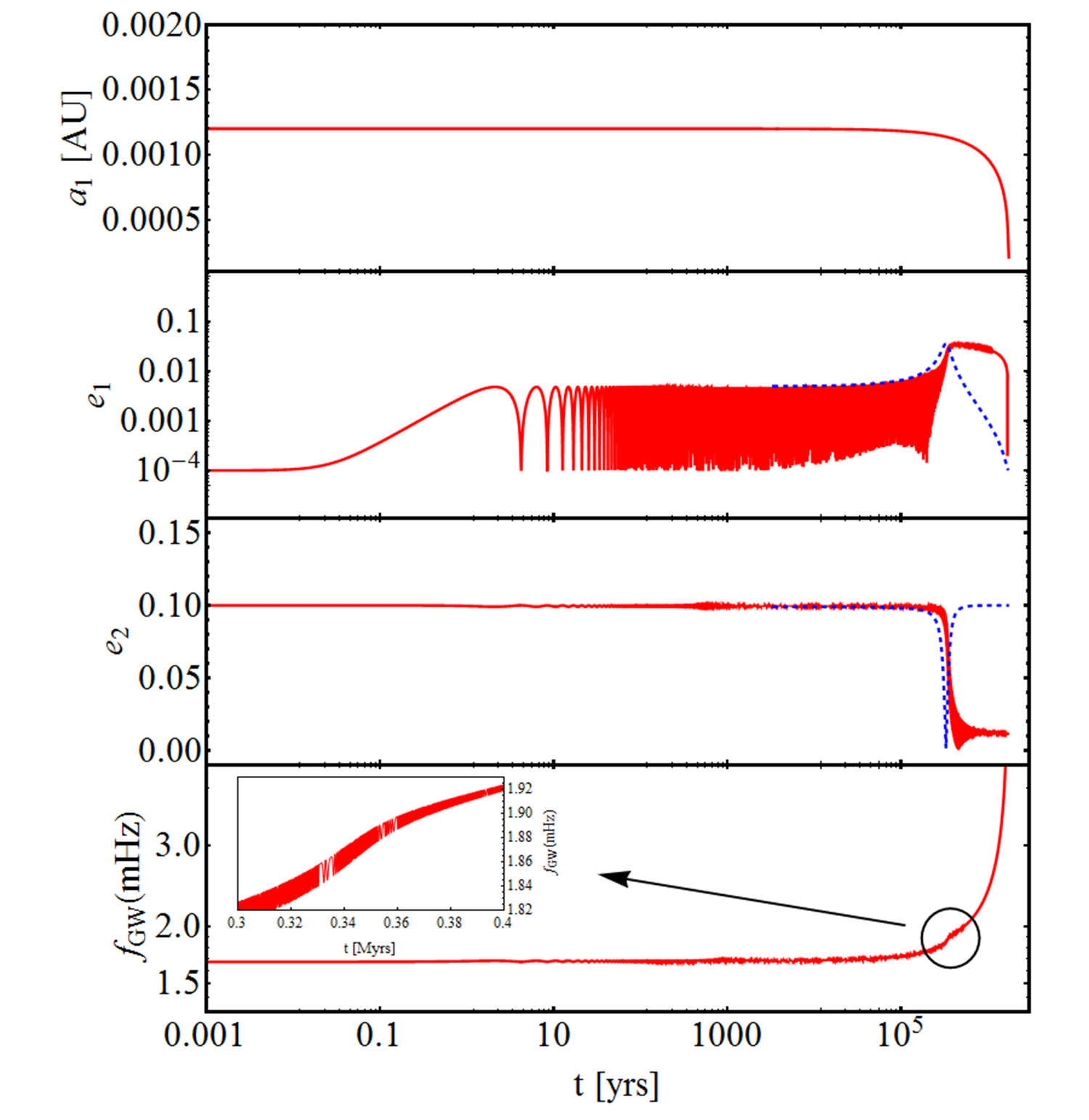}&
\includegraphics[width=8.5cm]{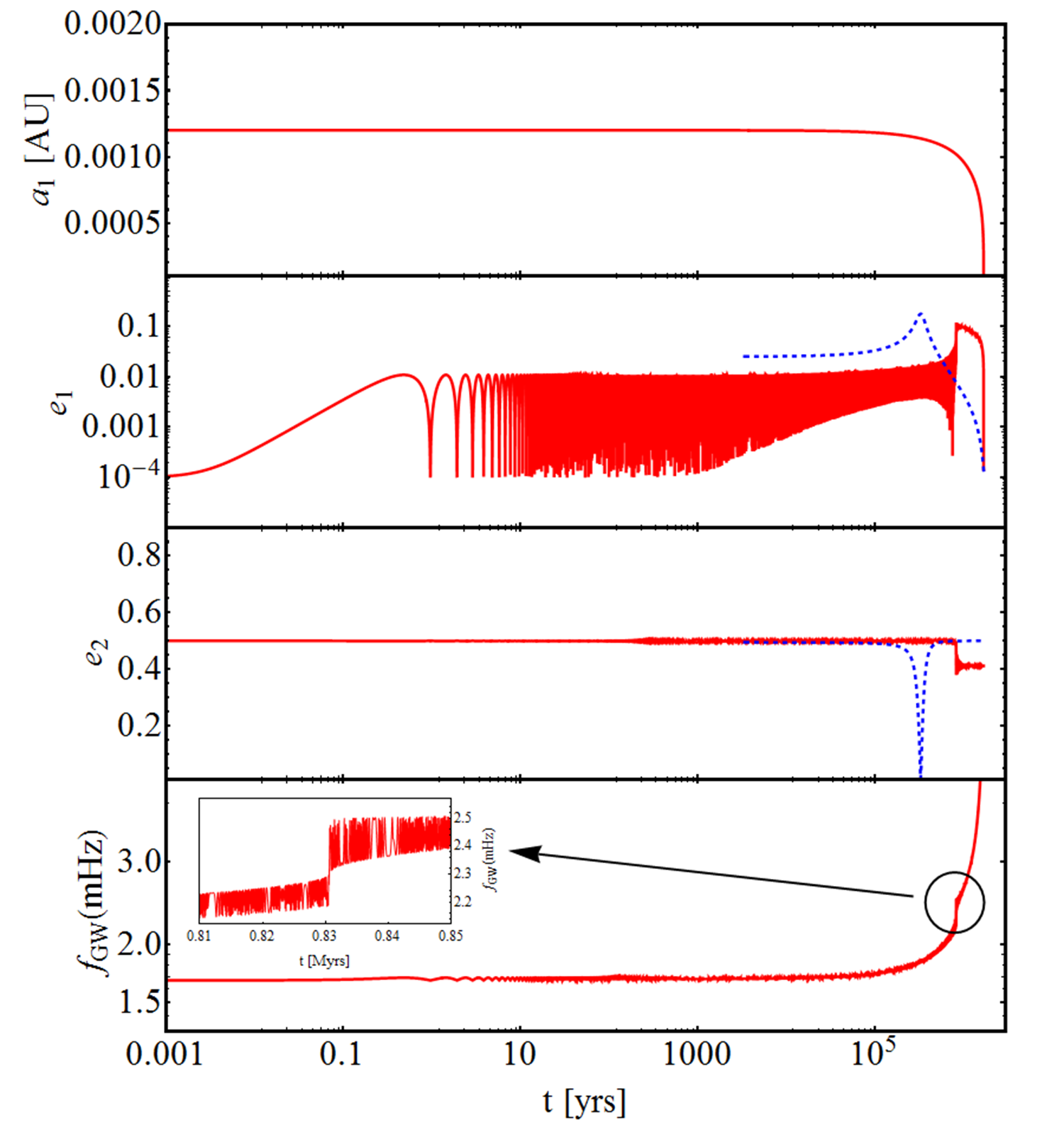} \\
\end{tabular}
\caption{The evolution of an inner merging WD-WD binary with a third
  body orbiting the mass center of the inner binary.  The masses of
  the WDs are $m_1=0.8\mathrm{M}_\odot$, $m_2=0.4\mathrm{M}_\odot$,
  respectively. The initial semimajor axis of the binary is
  $a_{1,0}=1.2\times10^{-3}\au$, the mass of the third body is
  $m_3=0.01\mathrm{M}_\odot$, the semimajor axis of the outer orbit
  is $a_{2}=1.27\times10^{-2}\au$.  The eccentricity of the inner
  binary and the longitude of the periapsis are initialized as
  $e_1=0.0001$, $\varpi_{1,0}=\varpi_{2,0}=0$, respectively.
  The outer binary is set up with two different initial eccentricities,
  $e_{2,0}=0.1$ for the left panel and $e_{2,0}=0.5$ for the right
  panel.  We numerically integrate Eqs.~(\ref{eq:C eqs1})-(\ref{eq:C eqs4}) with orbital decay [Eqs.\ref{eq:omegagr},
  \ref{eq:a1gw} and \ref{eq:e1gw}].  Different panels show: (top)
  the inner binary semimajor axis $a_1$ from numerical results (red) and
  analytical solution (dashed blue);
%DL: Above: WHAT IS THE ANALYTIC SOLUTION? IS IT NECESSARY?
 (top middle) the eccentricity of
  the inner binary $e_1$ from numerical integrations (red) and
  Eq.(\ref{eq:E1}) (dashed blue); (top bottom) the eccentricity of the
  outer binary $e_2$ from numerical integrations (red) and
  Eq.(\ref{eq:E2}) (dashed blue); and (bottom) the the peak
  GW frequency, also zoomed-in near the resonance.  Clearly,
  resonance occurs in both cases.
%DL:
% for both of the merging WD-WD
%  binary, and the behaviors become evident for large $e_{2,0}$.  Noted  that,
%DL: revised below
The analytical result for $e_{\rm lim}$ agrees with the numerical result
for small $e_{2,0}$.  }\label{fig:4}
\end{figure*}

In this section, we carry out calculations for the evolution of
coplanar systems, including the dissipative effect of gravitational
radiation. During the decay of the inner orbit, the system can
encounter the apsidal precession resonance for eccentricity excitation.
%DL:
However, it is not clear whether the peak eccentricity $e_{1,\peak}$
can be achieved as the system may pass through the resonance too quickly.
Also the analytical result in Sec. \ref{chap:Linear} is restricted to small eccentricities,
whereas the numerical calculations here apply to general eccentricities.

Due to gravitational radiation, the orbital semimajor axis of the
inner binary decays according to (e.g., \cite{Blaes})
\be\label{eq:a1gw}
\dot a_1|_\gw=-\frac{64}{5}\frac{G^3}{c^5}\frac{m_1 m_2(m_1+m_2)}{a_1^3(1-e_1^2)^{7/2}}\bigg(1+\frac{73}{24}e_1^2+\frac{37}{96}e_1^4\bigg).
\ee
Gravitational radiation also induces a secular change of $e_1$:
\be\label{eq:e1gw}
\dot e_1|_\gw=-\frac{304}{15}\frac{G^3}{c^5}\frac{m_1m_2(m_1+m_2)e_1}{a_1^4(1-\e1)^{5/2}}\bigg(1+\frac{121}{304}\e1\bigg).
\ee
%
%DL: delete
%and tends to promote periapse precession in Eq.(\ref{eq:C eqs1}).

Now consider an initially circular WD-WD binary
($m_1=0.8\mathrm{M}_\odot$, $m_2=0.4\mathrm{M}_\odot$) with the
semimajor axis $a_{1,0}=1.2\times10^{-3}\au$, and a small perturber
($m_3=0.01\mathrm{M}_\odot$) orbits the center mass of the inner
binary on a wider orbit ($a_{2}=1.27\times10^{-2}\au$).  We set up the
outer binary with two different initial eccentricities, $e_{2,0}=0.1$
and $e_{2,0}=0.5$.  Combining Eqs.(\ref{eq:omegagr}) and  (\ref{eq:a1gw})
with (\ref{eq:e1gw}), we integrate Eqs.~(\ref{eq:C eqs1})-(\ref{eq:C eqs4})
until the inner binary mergers.
%for the merging time scale ($a_{1}\rightarrow0\au$).
For each case, we also calculate the evolution of the peak GW
frequencies [Eq.(\ref{eq:crita1})].
%to explore the sensitivity of the resonant eccentricity.
The results are shown in Fig.\ref{fig:4}.

The left panel of Fig.\ref{fig:4} corresponds to the case of $e_{2,0}=0.1$.
%The system evolves as a function of
%time until the merger of inner orbit.
The eccentricity of the inner
binary undergoes small-amplitude oscillations at the early state, due
to the quadrupole potential from the outer binary.  During the gradual
orbital decay, $\dot\varpi_1$ becomes comparable with $\dot\varpi_2$,
the resonance starts to operate when $a_1$ is close to
$0.0011\au$ and forces $e_1$ to achieve $e_{1,\peak}$.  After that,
%DL:
gravitational radiation gradually reduces $e_1$, circularizing the inner binary
before the final merger.
%the GR effects become effective (in the high eccentricity phase) and
%lead the binary to be circularized before coalescence.
Corresponding to the excitation of $e_1$, the outer binary eccentricity $e_2$
decreases a lower value because of the resonance.
% and GWs frequency varies slightly.

%DL: I AM NOT SURE WHAT YOU MEAN BY ANALYTIC RESULT. HERE IS MY INTERPRETATION. PLS CHECK
To compare the numerical results (including gravitational radiation)
with the analytical solution presented in Sec. \ref{chap:Linear},
%of $e_1$ (Eq.\ref{eq:E1}) and $e_2$ (Eq.\ref{eq:E2}),
we calculate the maximum of $e_1$ and the minimum of $e_2$ attained for different values
of $a_1$ based on Eq.~(\ref{eq:E1}) and Eq.~(\ref{eq:E2}). The results are shown as
dashed curves in Fig.~\ref{fig:4}.
We see that the maximum $e_1$ attained in our numerical calculation agrees approximately
with the analytic $e_{1,{\rm peak}}$, with the agreement better for small $e_{2,0}$.
%DL: DELETE BELOW?
%It seems that the analytical solutions agree well with the numerical results
%except the time-maintenance in $e_{1,\peak}$.
%The former suffers a rapid decline while $e_{1,\peak}$ is hold for a long time in the latter.
%This may be the results from the evolving of $\varpi_1$ and $\varpi_2$.
%Noted that $\dot\varpi_{1,0}=\dot\varpi_{2,0}=0$ is keeping when we use Eq.(\ref{eq:E1})
%and Eq.(\ref{eq:E2}).
It is easy to notice that, in the numerical calculations, the time that
the system spends near $e_{1,\peak}$ is longer than the one in the analytical solution.
%DL:
Also, the peak $e_1$ is reached at somewhat smaller $a_1$
compared to analytical prediction; this difference arises from the finite value of
$e_{2,0}$.
Although this finite $e_{2,0}$ is unsatisfied with the linear approximation, it seems that the ``apsidal precession resonance"
still occurs and drives the eccentricity to the evident excitation,
as well as the GWs frequency.

%the peak from analytical result always appears a little bit
%earlier than the one in numerical integrations,
%which is associated with the finite value of $e_{2,0}$.

%DL:UNCLEAR. DELETE?
%The right panel of Fig.\ref{fig:3} shows the similar results but for
%the larger $e_{2,0}$.  The amplitude of oscillations in $e_1$ at
%resonance is relatively greater than the one in the above case.
%According to the evident excitation, GWs frequency varies a lot and
%the amplitude is enhanced by almost 0.2mHz within several decades.
%This maybe helpful for future detection in LISA.  However, the
%analytical solutions do not agree with numerical integrations here.

%DL:
%From this figure, we can conclude that although finite $e_{2,0}$ have
%broken the linear approximation, the ``apsidal precession resonance"
%still occur in the coplanar systems and the excited eccentricity is
%limited by $e_{2,0}$.

%DL: NOT DONE BELOW xxxxxxxxxxxxxxxxxxxxxxxxxxxxxxxxxxxxxxx
\section{Resonance in the inclined systems}
\label{chap:inclined}

\subsection{Without gravitational radiation}

\begin{figure*}
\centering
\begin{tabular}{cc}
\includegraphics[width=12cm]{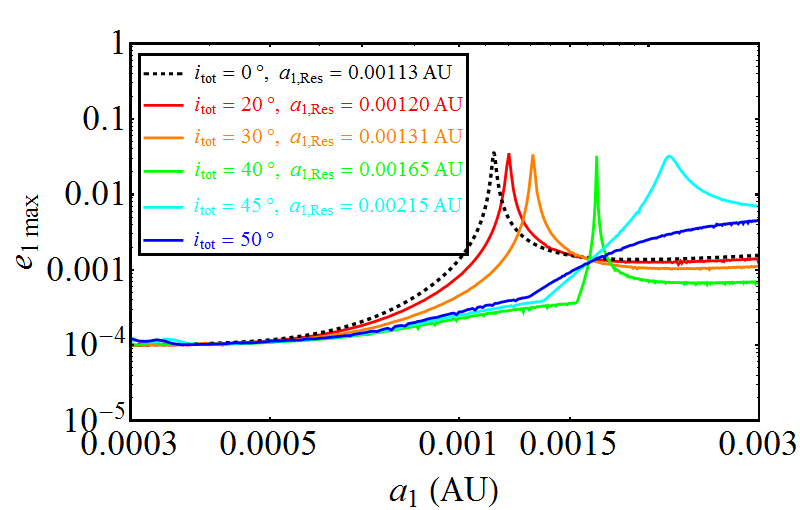}
\end{tabular}
%DL2: modified
\caption{The maximum eccentricity $e_{1,\m}$ as a function of $a_1$
  for triple systems with different initial mutual inclinations.  The
  triple system consists of a WD-WD binary ($m_1=0.8\mathrm{M}_\odot$,
  $m_2=0.4\mathrm{M}_\odot$) and a tertiary companion
  ($m_3=0.01\mathrm{M}_\odot$) with semimajor axis
  $a_{2}=1.27\times10^{-2}\au$.
%  For the each combination of initial
%  $\I$ and $a_{1,0}$, we integrate Eqs.(\ref{eq:omegagr}) and
%  (A1)-(A9) in \cite{SRF} for $\sim2t_K/\varepsilon_\oct$ and record
%  $e_{1,\m}$.  For small inclinations
  ($i_{\mathrm{tot},0}<50^\circ$), there exists a distinct peak in
  $e_{1,\m}$, while for large inclinations
  ($i_{\mathrm{tot},0}\geq50^\circ$), the peak is erased by
  Lidov-Kozai oscillations.  We denote $a_{1,\mathrm{Res}}$ the
  location where $e_{1,\peak}$ is achieved.  }\label{fig:5}
\end{figure*}

%DL2:
%In the coplanar limit, the occurrence of resonance depends on the mass
%and location of the third body. It is natural to justify whether the
%resonance is associated with the inclination angles.
In this section, we extend our calculations to the general cases of
mutually inclined inner/outer binaries.  Since no simple analytical result
can be derived, we sample $a_1$ to determine the resonance location
numerically for given outer binary parameters.

As an example, we consider a triple system consisting of a WD-WD inner
binary ($m_1=0.8\mathrm{M}_\odot$, $m_2=0.4\mathrm{M}_\odot$) and a
tertiary companion ($m_3=0.01\mathrm{M}_\odot$) with semimajor
axis $a_{2}=1.27\times10^{-2}\au$.  The initial eccentricities of the two
orbits are $e_{1,0}=0.0001$ and $e_{2,0}=0.1$.  We set up the system
by varying the initial semimajor axis $a_{1,0}$ between $0.0003\au$
and $0.003\au$, and $i_{\mathrm{tot},0}$ between $0^\circ$ and
$50^\circ$.  For the each combination of the initial $a_{1,0}$ and
$i_{\mathrm{tot},0}$, we integrate Eqs.(\ref{eq:omegagr}) and
(A1)-(A9) in \cite{SRF} [without evolving Eq.(\ref{eq:a1gw})] for a
duration of $\sim2t_K/\varepsilon_\oct$, and record the maximum
eccentricity $e_{1,\m}$ attained during the evolution.  The results
are shown in Fig.\ref{fig:5}.  We see that for $\I<50^\circ$, there
exists a distinct peak in $e_{1,\m}$ at the location
$a_1=a_{1,\mathrm{Res}}$. This resonant location increases with increasing $\I$.
%Noted that for the coplanar case (dashed curve), the result
%is equivalent to the analytical solutions (Eqs.\ref{eq:E1}-\ref{eq:E2}).
%Here, we introduce the parameter as $a_{1,\mathrm{Res}}$ to denote the location,
%where $e_{1,\peak}$ is achieved.
%DL2: I have deleted below: I don't see ``oscillation'' in Fig.4??
%\textbf{As shown in Fig.\ref{fig:4},}
%in the region of $a_1>a_{1,\mathrm{Res}}$,
%the quadrupole potential induced by the third body
%drive $e_1$ to undergo oscillations,
%somewhere the Lidov-Kozai mechanism become efficient for the relative larger $\I$.
%\textbf{With the decrease of the initial values of $a_1$,
%the GR effects tend to play an important role,
%driving the eccentricity to reach the peak value.
%In the region of $a_1<a_{1,\mathrm{Res}}$,
%$e_{1,\m}$ gets a rapid decline.
%This corresponds to the resonant and circularized state in the orbital
%evolution (see section \ref{chap:inclined GW}).
Interestingly, the value of $e_{1,\peak}$ is almost independent of
the inclination angle. For $\I> 50^\circ$, the peak
in the $e_{1,{\rm max}}$-$a_1$ curve becomes broader as the
inner binary begins to experience Lidov-Kozai oscillations.

\begin{figure}
\centering
\begin{tabular}{cc}
\includegraphics[width=8cm]{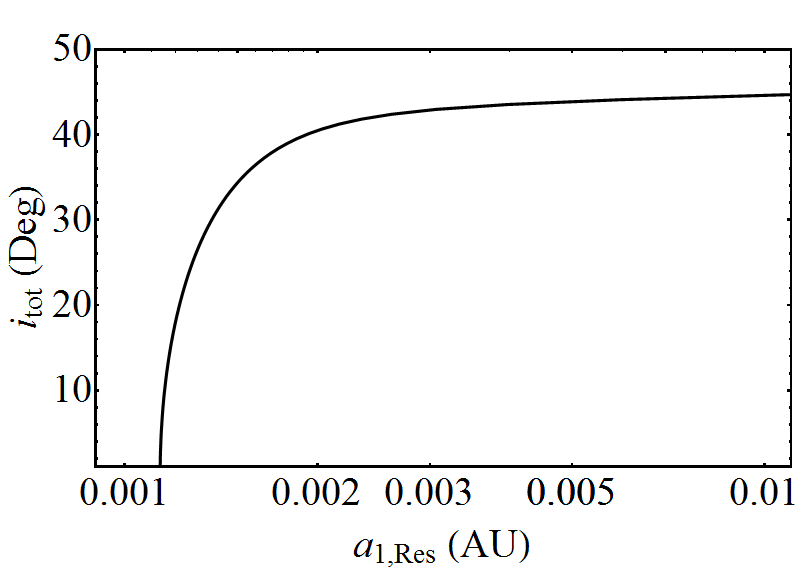}
\end{tabular}
%DL2: modified
\caption{The relationship between the inclination $\I$ and $a_{1,\mathrm{Res}}$
in the case which is shown in Fig.\ref{fig:5}.
This result is from Eqs.(\ref{eq:inclinedomega1})-(\ref{eq:inclinedomega2}).
}\label{fig:6}
\end{figure}

As discussed in Sec. \ref{chap:Linear}, the apsidal precession resonance occurs
when $\dot\varpi_1\sim\dot\varpi_2$. For inclined systems,
%It seems that the resonance in the inclined system is still responsible for this excitation.
we have (see Eqs.(A6)-(A7) at the quadrupole level in \cite{SRF})
%DL2: Eq.42 and Eq.43 should reduce to eq.21 and 22?? Please check
\be\label{eq:inclinedomega1}
\dot\omega_1=\frac{3}{4}\bigg(\frac{a_1}{a_2}\bigg)^3n_1\frac{m_3}{m_1+m_2}\big(2-\cos\I\big)
+\dot\omega_\gr|_{e_1\rightarrow0},
\ee
\be\label{eq:inclinedomega2}
\dot\omega_2=\frac{3}{16}\bigg(\frac{a_1}{a_2}\bigg)^2n_2\frac{m_1m_2}{(m_1+m_2)^2}\big(3-4\cos\I+5\cos2\I\big),
\ee
where we have set $e_1=e_2=\omega_1=0$.
%DL2: Pls rewrite the following.  Maybe a plot is useful? a_1,res as a function of i_tot??
%DL3: Add. Pls check
Although we do not have an exact resonance condition for inclined
systems, it is reasonable to assume that $\dot\omega_1=\dot\omega_2$
provides a good estimate to the resonance radius.  Equating
Eq.~(\ref{eq:inclinedomega1}) and Eq.~(\ref{eq:inclinedomega2}), we
obtain $a_{1,\mathrm{Res}}$ as a function of $\I$. The result is shown
in Fig.\ref{fig:6}. We see that, although this analytic resonant
radius does not exactly agree with the numerical value depicted in
Fig.\ref{fig:5}, especially for $\I\gtrsim 40^\circ$, it nevertheless
reproduces the general feature of Fig.\ref{fig:5}. In particular, we see that
$a_{1,\mathrm{Res}}$ shifts to larger values as $\I$ increases, and
the resonance disappears when $\I\gtrsim 45^\circ$.
Thus,  the apsidal precession resonance condition $\dot\omega_1=\dot\omega_2$
provides a good description for eccentricity excitation in inclined systems.

\subsection{With gravitational radiation}
\label{chap:inclined GW}

\begin{figure*}
\centering
\begin{tabular}{cc}
\includegraphics[width=9cm]{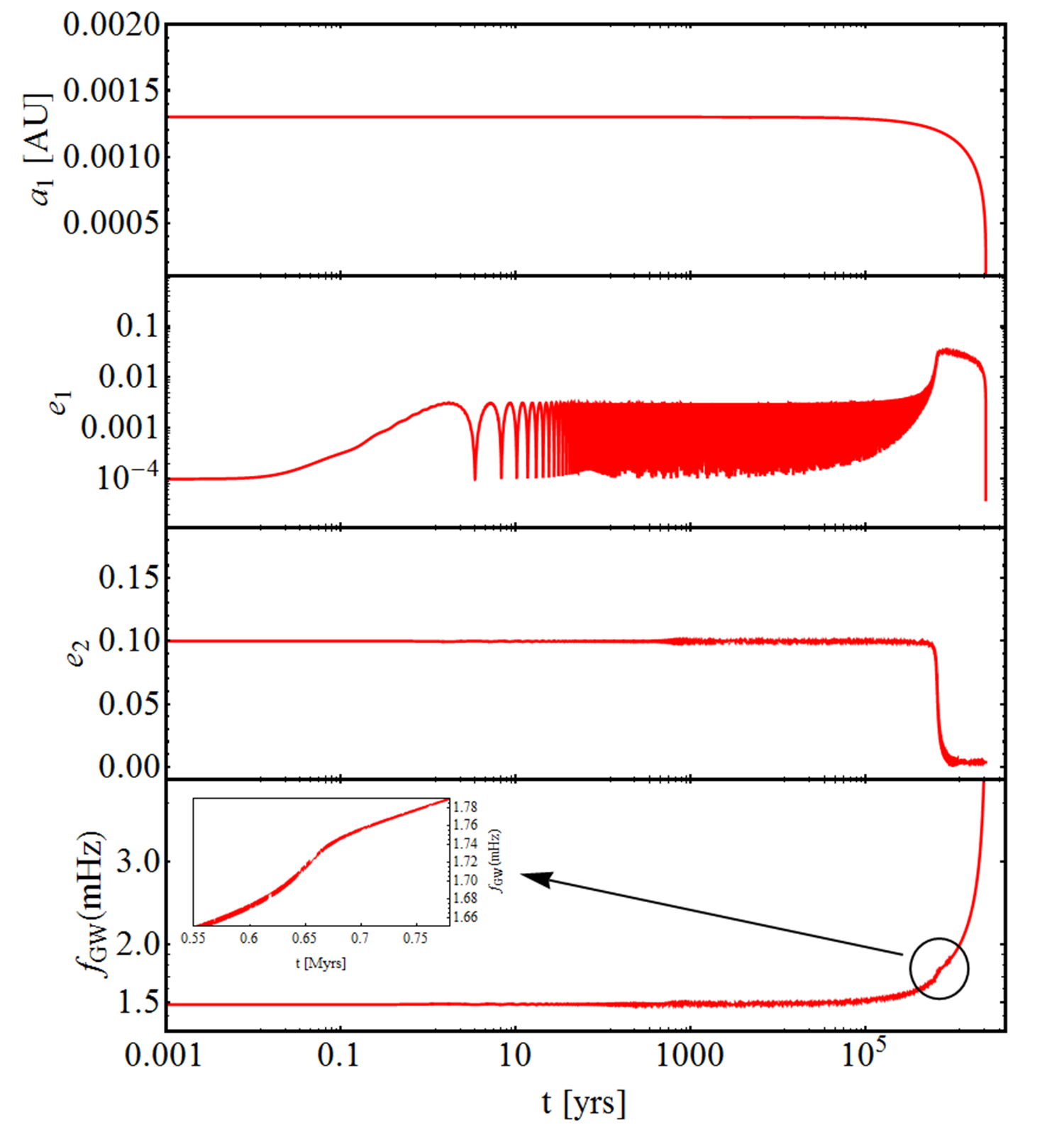}&
\includegraphics[width=9cm]{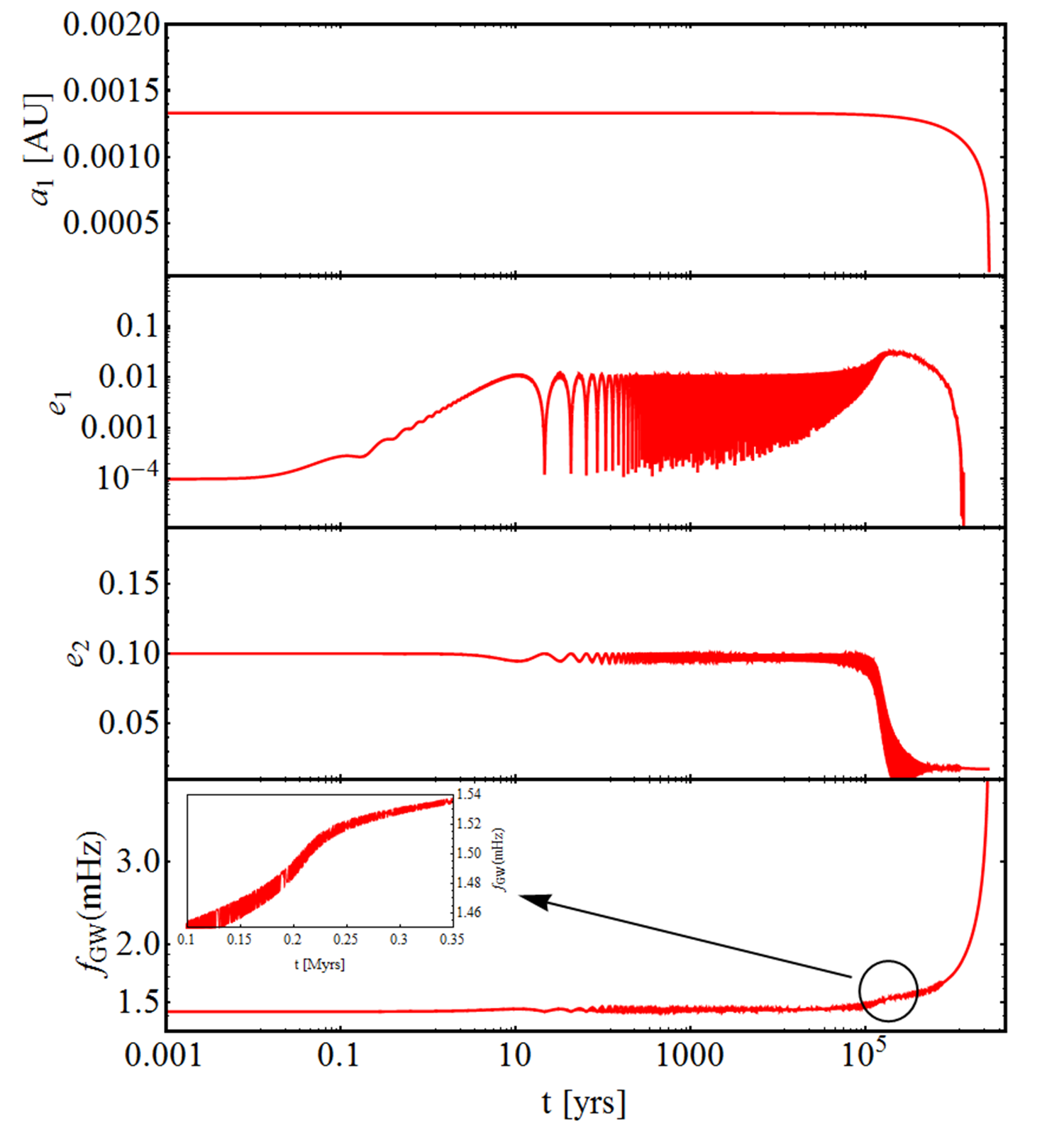} \\
\end{tabular}
%DL2:
\caption{Similar to Fig.~\ref{fig:4}, but for triple systems with
  finite initial mutual inclinations.  The stellar masses and initial
  eccentricities the same as in the left panel of Fig.~\ref{fig:4}.
  Two examples are shown: In the left panels, we choose
  $i_{\mathrm{tot},0}=20^\circ$ and $a_{1,0}=1.30\times10^{-3}\au$;
%  corresponds to an inner and outer inclinations with respect to the
%  total angular momentum of $i_{1,0}=2.05^\circ$ and
%  $i_{2,0}=17.95^\circ$;
in the right panel, $i_{\mathrm{tot},0}=30^\circ$ and
  $a_{1,0}=1.35\times10^{-3}\au$.
% so that $i_{1,0}=3^\circ$ and $i_{2,0}=27^\circ$.
%Here, we integrate the full Eqs.(A1)-(A9) in
%  \cite{SRF} numerically combining Eqs.(\ref{eq:omegagr}),
%  (\ref{eq:a1gw}) with (\ref{eq:e1gw}).
The qualitative behavior of the inclined systems
is similar to the coplanar system (see Fig.\ref{fig:4}).
% the excitation still occur for non-sufficient large inclined systems.  But the
%  amplitude of the oscillations in $e_1$ before the resonance is
%  relative larger than the coplanar case and $a_{1,\mathrm{Res}}$ is
%  shifted slightly.
}\label{fig:7}
\end{figure*}

As in the Sec. \ref{chap:coplanar GW}, we now include gravitational
radiation in the secular equations for inclined triples
[Eqs.~(A1)-(A9) in \cite{SRF} with Eqs.~(\ref{eq:omegagr}),
(\ref{eq:a1gw}) and (\ref{eq:e1gw})], numerically evolving the
system through the apsidal precession resonance.

%we explore the merging binaries go through the resonant state by the
%numerical integrations involving GWs radiation,
%this time for triple system in inclined configurations.
%Note that, the Lidov-Kozai oscillations become
%important if the relative inclination is sufficient large, and the
%excitation of the inner eccentricity may be erased.  So, the initial
%$\I$ studied here are set to be several small angles.

% DL2: BIN, PLS LOOK AT THE FIRST SENTENCE BELOW: IT HAS 2 VERBS (...EXPLORE... GO...).
%As shown in the section \ref{chap:coplanar GW},
%we explore the merging binaries go through the resonant state by the
%numerical integrations involving GWs radiation,
%this time for triple system in inclined configurations.
%Note that, the Lidov-Kozai oscillations become
%important if the relative inclination is sufficient large, and the
%excitation of the inner eccentricity may be erased.  So, the initial
%$\I$ studied here are set to be several small angles.

We consider two examples with $i_{\mathrm{tot},0}=20^\circ$ and
$30^\circ$.  Since $a_{1,\mathrm{Res}}$ is shifted to a larger value
compared to the $i_{\mathrm{tot},0}=0$ case, we set the initial
semimajor axis as $a_{1,0}=1.30\times10^{-3}\au$ for
$i_{\mathrm{tot},0}=20^\circ$ and $a_{1,0}=1.35\times10^{-3}\au$ for
$i_{\mathrm{tot},0}=30^\circ$.  Note that the initial mutual
inclination of $20^\circ$ corresponds to an inner and outer
inclinations with respect to the total angular momentum of
$2.05^\circ$ and $17.95^\circ$.  Similarly, for
$i_{\mathrm{tot},0}=30^\circ$ we have $i_{1,0}=3^\circ$ and
$i_{2,0}=27^\circ$. We also set the initial arguments of pericenter
($\omega_{1,0}$, $\omega_{2,0}$) and the ascending node ($\Omega_{1,0}$ and $\Omega_{2,0}$)
of the inner and outer binaries to zero.
The results are shown in Fig.~\ref{fig:7}.
We find that the behaviors of $e_1$ and $e_2$, in particular the
excitation of the inner binary eccentricity, are
similar to the coplanar case.
When the resonance is encountered during the orbital decay,
$e_1$ increases while $e_2$ decreases.
The peak eccentricity $e_{1,\peak}$ has a similar value as the
coplanar case.
The $e_1$ oscillation has a larger amplitude
before the resonance for the higher $i_{\mathrm{tot},0}$.

For systems with initial inclinations
$i_{\mathrm{tot},0}\geq50^\circ$, we find that the inner binary does
not undergo a distinct resonant eccentricity excitation. Instead, it
experiences a gentle and modest eccentricity growth, followed by
circularization directly due to GW emission.

%From this figure, we find the excitation still occur for non-sufficient large inclined systems,
%although the resonant condition cannot be understood analytically.
%Comparing with the coplanar case, the inclination angle does not change the value of $e_{1,\peak}$
%but shift $a_{1,\mathrm{Res}}$ slightly.

%DL3: I have rewritten this section
\section{Summary and Discussion}

In this paper we have studied the effect of an external tertiary
companion on compact binaries (consisting of WDs, NSs or BHs)
undergoing orbital decay due to gravitational radiation.  We find
that, during the orbital decay, the binary eccentricity can be excited
due to gravitational perturbation from the tertiary.  The eccentricity
excitation occurs when the system passes through an ``apsidal
precession resonance", when the pericenter precession rate of the
inner binary, due to the combined effects of gravitational
perturbation of the external companion and general relativity, equals
the precession rate of the outer binary.  This eccentricity excitation
can be thought of as ``eccentricity exchange'' between the inner and
outer binaries. It requires that the outer companion be on an
eccentric orbit. It also requires that the mutual inclination between
the inner and outer orbits be less than $\sim 40^\circ$, so that the
resonance is not erased by Lidov-Kozai oscillations.  Thus, as the
triple system passes through the resonance (as a result of orbital
decay of the inner binary), the outer binary ``gives'' part or all of
its eccentricity to the inner binary.

Through analytical calculations (for coplanar systems with small
eccentricities) and numerical integrations of the secular evolution
equations, we have carried out a detailed analysis of the apsidal
precession resonance and more importantly, we have examined the
behavior of the system during resonance passage.  For example, the
value of the excited eccentricity, $e_{1,\peak}$, can be estimated by
Eq.(\ref{eq:peak}) and depends on the mass of the third body.  In
general, as the mass increases, $e_{1,\peak}$ becomes larger, but
cannot exceed the initial eccentricity of the outer orbit (see
Fig.\ref{fig:3}).

Obviously, the resonant excitation/exchange of eccentricity during
binary orbital decay can significantly influence the gravitational
wave signals from the binary. We have identified the parameter space
where the resonance occurs. For some system parameters (e.g., a WD
binary with a brown dwarf tertiary), the resonance can happen when the
binary emits GWs in the $10^{-4}-10^{-1}$~Hz range (the sensitivity
band of LISA). Our results indicate that the excitation of
eccentricity is possible for a variety of compact binaries if there
exists an appropriate tertiary companion that satisfies the resonant
condition.

\section*{Acknowledgments}
%DL3:
Bin Liu thanks Diego J. Mu\~noz for some helpful discussion.  This
work is supported in part by grants from the 
National Basic Research Program of China
(No. 2012CB821801), the Strategic Priority Research Program of the Chinese
Academy of Sciences (No. XDB09000000), and the National Natural Science 
Foundation (No. U1431228, No. 11133005, No. 11233003, No. 11421303).  
It is also supported in part by National Science Foundation Grant
No. AST-1211061 and NASA Grants No. NNX14AG94G and No. NNX14AP31G.


\begin{thebibliography}{99}

\bibitem[\protect\citeauthoryear{Webbink}{1984}]{Webbink} R.~F. Webbink, Astrophys. J. \textbf{277}, 355 (1984).

\bibitem[\protect\citeauthoryear{I. Iben, Jr. and Tutukov}{1984}]{IT} I. Iben, Jr. and A.~V. Tutukov, Astrophys. J. Suppl. Ser.
\textbf{54}, 335 (1984).

\bibitem[\protect\citeauthoryear{Eichler et
al.}{1989}]{Eichler} D. Eichler, M. Livio, T. Piran, and D.~N. Schramm, Nature (London) \textbf{340}, 126 (1989).

\bibitem[\protect\citeauthoryear{Berger}{2011}]{Berger} E. Berger, New Astron. Rev. \textbf{55}, 1 (2011).

\bibitem[\protect\citeauthoryear{Nissanke, Kasliwal, and Georgieva}{2013}]{Nissanke} S. Nissanke, M. Kasliwal, and A. Georgieva, Astrophys. J. \textbf{767}, 124 (2013).

\bibitem[\protect\citeauthoryear{Nelemans}{2009}]{Nelemans} G. Nelemans, Classical Quantum Gravity \textbf{26}, 094030 (2009).

\bibitem[\protect\citeauthoryear{elisascience}{}]{elisascience} eLISA Gravitational Wave Observatory, 
https://www.elisascience.org/.

\bibitem[\protect\citeauthoryear{LIGO}{}]{LIGO} LIGO Scientific Collaboration, http://www.ligo.org/.

\bibitem[\protect\citeauthoryear{Cutler et al.}{1993}]{Cutler} C. Cutler, T. A. Apostolatos, L. Bildsten, L. S. Finn, E. E. Flanagan, D. Kennefick, D. M. Markovic, A. Ori, E. Poisson, G. J. Sussman, and K. S. Thorne, Phys. Rev. Lett. \textbf{70}, 2984 (1993).

\bibitem[\protect\citeauthoryear{M. Shibata, and K. Taniguchi}{2006}]{Shibata} M. Shibata, and K. Taniguchi, Phys. Rev. D \textbf{73}, 064027 (2006).

\bibitem[\protect\citeauthoryear{F. Foucart et al.}{2012}]{Foucart} F. Foucart, M. D. Duez, L. E. Kidder, M. A. Scheel, B. Szilagyi, and S. A. Teukolsky, Phys. Rev. D \textbf{85}, 044015 (2012).

\bibitem[\protect\citeauthoryear{Y. Sekiguchi et al.}{2012}]{Sekiguchi} Y. Sekiguchi, K. Kiuchi, K. Kyutoku, and M. Shibata, Prog. Theor. Exp. Phys. (\textbf{2012}), 01A304.

\bibitem[\protect\citeauthoryear{Tokovinin}{2014}]{Tokovinin} A. Tokovinin, Astron. J. \textbf{147}, 87 (2014).


%\bibitem[\protect\citeauthoryear{Bildsten and Cutler}{1992}]{Bildsten} L. Bildsten, and C. Cutler, Astrophys. J. \textbf{400}, 175 (1992).
%
%\bibitem[\protect\citeauthoryear{Lai and Wiseman}{1996}]{LaiW} D. Lai, and A.~G. Wiseman,  Phys. Rev. D \textbf{54}, 3958 (1996).
%
%\bibitem[\protect\citeauthoryear{Wiggins and Lai}{2000}]{WLai} P. Wiggins, and D. Lai, Astrophys. J. \textbf{532}, 530 (2000).

\bibitem[\protect\citeauthoryear{Lidov}{1962}]{Lidov} M. L. Lidov, Planet. Space Sci. \textbf{9}, 719 (1962).

\bibitem[\protect\citeauthoryear{Kozai}{1962}]{Kozai} Y. Kozai, Astron. J. \textbf{67}, 591 (1962).

\bibitem[\protect\citeauthoryear{Eggleton and Kiseleva\---Eggleton}{2001}]{Eggleton} P. P. Eggleton and L. Kiseleva-Eggleton, Astrophys. J. \textbf{562}, 1012 (2001).

\bibitem[\protect\citeauthoryear{Fabrycky and Tremaine}{2007}]{FT} D. C. Fabrycky and S. Tremaine, Astrophys. J. \textbf{669} 1298 (2007).

\bibitem[\protect\citeauthoryear{Holman, Touma, and Tremaine}{1997}]{Holman} M. Holman, J. Touma, and S. Tremain, Nature (London) \textbf{386}, 254 (1997).

\bibitem[\protect\citeauthoryear{Innanen et al.}{1997}]{Innanen} K. A. Innanen, J. Q. Zheng, S. Mikkola, and M. J. Valtonen, Astron. J. \textbf{113}, 1915 (1997).

\bibitem[\protect\citeauthoryear{Wu and Murray}{2003}]{WM} Y. Wu, and N. Murray, Astrophys. J. \textbf{589}, 605 (2003).

\bibitem[\protect\citeauthoryear{Naoz et al.}{2011}]{Smadar 2011} S. Naoz, W. M. Farr, Y. Lithwick, F. A. Rasio, and J. Teyssandier, Nature (London) \textbf{473}, 187 (2011).

\bibitem[\protect\citeauthoryear{Thompson}{2011}]{Thompson} T.~A. Thompson, Astrophys. J. \textbf{741}, 82 (2011).

\bibitem[\protect\citeauthoryear{Prodan, Murray, and Thompson}{2013}]{PMT} S. Prodan, N. Murray, and T.~A. Thompson, arXiv:1305.2191.

\bibitem[\protect\citeauthoryear{Katz, Dong, and Malhotra}{2011}]{Katz PRL} B. Katz, S. Dong, and R. Malhotra, Phys. Rev. Lett. \textbf{107}, 181101 (2011)

\bibitem[\protect\citeauthoryear{Blaes et al.}{2002}]{Blaes} O. Blaes, M. H. Lee, and A. Socrates, Astrophys. J. \textbf{578}, 775 (2002).

\bibitem[\protect\citeauthoryear{Miller and Hamilton}{2002}]{MH} M. C. Miller and D.P. Hamilton, Astrophys. J. \textbf{576}, 894 (2002).

\bibitem[\protect\citeauthoryear{Wen}{2003}]{Wen} L. Wen, Astrophys. J. \textbf{598}, 419 (2003).

\bibitem[\protect\citeauthoryear{Antonini, Murray and Mikkola}{2014}]{AMM} F. Antonini, N. Murray, and S. Mikkola, Astrophys. J. \textbf{781}, 45 (2014).

\bibitem[\protect\citeauthoryear{Storch, Anderson,
\& Lai}{2014}]{DongLai1} N.~I. Storch, K.~R. Anderson, and D. Lai, Science \textbf{345}, 1317 (2014)

\bibitem[\protect\citeauthoryear{Storch \& Lai}{2015}]{DongLai2} N.~I. Storch and D. Lai, Mon. Not. R. Astr. Soc. \textbf{448}, 1821 (2015).

\bibitem[\protect\citeauthoryear{Harrington}{1968}]{Harrington} R. S. Harrington, Astron. J. \textbf{73}, 190 (1968).

\bibitem[\protect\citeauthoryear{Marchal et al.}{1990}]{Marchal} C. Marchal, {\it The Three-Body Problem} (Elsevier, Amsterdam, 1990).

\bibitem[\protect\citeauthoryear{Ford et al.}{2000b}]{Ford} E. B. Ford, B. Kozinsky, and F. A. Rasio, Astrophys. J. \textbf{535}, 385 (2000b).

\bibitem[\protect\citeauthoryear{Naoz et al.}{2013b}]{Smadar 2013b} S. Naoz, W. M. Farr, Y. Lithwick, and F. A. Rasio, Mon. Not. R. Astr. Soc. \textbf{431}, 2155 (2013).

\bibitem[\protect\citeauthoryear{Liu, Mu{\~n}oz,
and Lai}{2015}]{SRF} B. Liu, D.~J. Mu{\~n}oz, and D. Lai, Mon. Not. R. Astr. Soc. \textbf{447}, 747 (2015).

\bibitem[\protect\citeauthoryear{Mardling}{2007}]{Mardling 2007} R.~A. Mardling, Mon. Not. R. Astr. Soc. \textbf{382}, 1768 (2007).

\bibitem[\protect\citeauthoryear{Naoz et al.}{2013}]{Smadar 2013a} S. Naoz, B. Kocsis, A. Loeb, and N. Yunes, Astrophys. J. \textbf{773}, 187 (2013).

\bibitem[\protect\citeauthoryear{Murray and Dermott}{1999}]{MD} C. D. Murray and S. F. Dermott, 
{\it Solar System Dynamics} (Cambridge University Press, Cambridge, England, 1999).

\bibitem[\protect\citeauthoryear{Wu and Goldreich}{2002}]{Wu G} Y. Wu and P. Goldreich, Astrophys. J. \textbf{564}, 1024 (2002).

\bibitem[\protect\citeauthoryear{Mardling and Aarseth}{2001}]{MA} R. A. Mardling and S. J. Aarseth, Mon. Not. R. Astr. Soc. \textbf{321}, 398 (2001).

\bibitem[\protect\citeauthoryear{Peters}{1964}]{peters} P. C. Peters, Phys. Rev. \textbf{136}, 1224 (1964).

%\bibitem[\protect\citeauthoryear{Mardling}{2010}]{Mardling 2010} R.~A. Mardling, Mon. Not. R. Astr. Soc. \textbf{407}, 1048 (2010).










\end{thebibliography}
\end{document}